\documentclass[runningheads]{llncs}
\usepackage[T1]{fontenc}
\usepackage{graphicx}
\usepackage{subfigure}
\usepackage{makecell}
\usepackage{hyperref}
\usepackage{amsfonts}
\usepackage{anyfontsize}
\usepackage[misc]{ifsym}

\begin{document}
\title{EDUE: Expert Disagreement-Guided One-Pass Uncertainty Estimation for Medical Image Segmentation}
% XD-UE
\author{Kudaibergen Abutalip\inst{1}\textsuperscript{\Letter} \and
Numan Saeed\inst{1} \and
Ikboljon Sobirov\inst{1} \and
Vincent Andrearczyk\inst{2} \and
Adrien Depeursinge\inst{2} \and
Mohammad Yaqub\inst{1}
}
\institute{
Mohamed bin Zayed University of Artificial Intelligence, Abu Dhabi, UAE \\ \href{mailto:kudaibergen.abutalip@mbzuai.ac.ae}{\texttt{kudaibergen.abutalip@mbzuai.ac.ae}} \and
University of Applied Sciences Western Switzerland (HES-SO), Sierre, Switzerland}
\maketitle              % typeset the header of the contribution
\begin{abstract}
 Deploying deep learning (DL) models in medical applications relies on predictive performance and other critical factors, such as conveying trustworthy predictive uncertainty. Uncertainty estimation (UE) methods provide potential solutions for evaluating prediction reliability and improving the model confidence calibration. Despite increasing interest in UE, challenges persist, such as the need for explicit methods to capture aleatoric uncertainty and align uncertainty estimates with real-life disagreements among domain experts. This paper proposes an Expert Disagreement-Guided Uncertainty Estimation (EDUE) for medical image segmentation. By leveraging variability in ground-truth annotations from multiple raters, we guide the model during training and incorporate random sampling-based strategies to enhance calibration confidence. Our method achieves $55\%$ and $23\%$ improvement in correlation on average with expert disagreements at the image and pixel levels, respectively, better calibration, and competitive segmentation performance compared to the state-of-the-art deep ensembles, requiring only a single forward pass.

 % demonstrates well-calibrated uncertainty outputs with improved correlation with expert disagreements compared to existing methods, 
\end{abstract}

\section{Introduction}
Maximizing the predictive performance of deep learning (DL) models is not the only factor leading to a wide-scale deployment in real-world applications. Particularly in the medical domain, many other model properties must be analyzed to ensure clinical adoption and minimize unforeseen consequences. Experts underline the inability of models to convey trustworthy predictive uncertainty as one of the main reasons for their slow and limited adoption in clinical practice~\cite{banerji2023clinical}. Uncertainty estimation (UE) is gaining attention as one of the potential solutions for evaluating prediction reliability as well as for purposes such as enhancing prediction quality, conducting quality assurance, domain adaptation, and active learning \cite{Gawlikowski2023,ZOU2023100003}. 

Various UE methods have been proposed in the literature. Stochastic Variational Inference (SVI)\cite{Blundell2015WeightUI} estimates the posterior distribution by modeling a Gaussian distribution for each parameter of the network. Monte-Carlo Dropout (MCDO) \cite{pmlr-v48-gal16} aggregates outputs of multiple forward passes of the same input with activated dropout layers to approximate the true posterior of the model. Deep Ensembles (DE) \cite{NIPS2017_9ef2ed4b} consists of multiple networks trained with different initializations. Test-time data augmentation (TTA) \cite{Ayhan2018TesttimeDA} also uses multiple forward passes but with differently augmented versions of the same input. Studies \cite{Mehrtens2023BenchmarkingCU,Ashukha2020PitfallsOI} highlight that DE outperforms most methods in robustness and confidence calibration despite their time and memory inefficiency. Although a growing number of studies on UE indicate a promising trajectory, some questions remain unanswered. 

A recent study~\cite{kahl2024values} underscored several pitfalls of UE works. As one of the main recommendations, the authors suggest developing an explicit means of assessment for capturing aleatoric uncertainty by having references from multiple annotators that reflect variability in the data. This highlights situations when there is a discrepancy between uncertainty estimates and disagreements among domain experts. Inter- and intra-observer variability is a recognized challenge in the process of annotating medical images, as experts often have different opinions and levels of expertise when assigning labels \cite{schaekermann2019understanding,joskowicz2019inter,kumar2007glaucoma}. To illustrate, Tables A2 and A3 in the Appendix show the inter-observer agreement scores for the retinal fundus images for glaucoma analysis (RIGA) \cite{Almazroa2017-cp} and head and neck tumor segmentation (HECKTOR) \cite{ANDREARCZYK2023102972} datasets used in this study. Although multi-rater label sampling and training strategies exist \cite{ji2021learning,Liao2023TAB}, the direction of explicitly incorporating this natural uncertainty information into the training process to obtain better calibrated and more reliable model outputs remains underexplored. 

As models are trained to mimic human annotators for disease detection, there is a need to align the uncertainty estimates with real-life divergences in the opinions of annotators for different scenarios. If we focus on trust and transparency, such an alignment can foster more effective collaboration between humans and DL models. When models express uncertainty in situations that mirror human uncertainty, users are more inclined to trust the model's predictions. Moreover, enhancing the ability to handle scenarios deviating from training data promotes robustness and adaptability.

% Similarly to how we train models to imitate human annotators to detect diseases automatically, we might also want to train them to have similar uncertainty estimates that reflect real-life divergences in the opinions of annotators. This necessitates the development of more comprehensive UE approaches. 

Additionally, the development of new UE methods should emphasize simplicity and efficiency to ensure widespread adoption and accessibility. Most of the current UE methods require several input passes (e.g., MCDO, TTA) or considerably increase the number of parameters, which incurs additional time and financial and environmental costs.

To address these issues, we propose an expert disagreement-guided uncertainty estimation (EDUE) method for medical image segmentation in this study. We explicitly use variability in ground-truth annotations from several raters to guide the model during training. In addition, previous studies~\cite{jensen2019improving,jungo2018effect} have highlighted the benefits of using multi-rater labels for improved calibration. Accordingly, we employ a random sampling-based strategy to incorporate ground-truth masks from all annotators during training. We validate our results on two distinct ophthalmology and head and neck tumor datasets and show that EDUE produces well-calibrated uncertainty outputs that correlate better with expert disagreements than existing state-of-the-art methods. The main contributions of this study are as follows:
\begin{itemize}
    \item We develop a novel, simple and intuitive UE method that takes into account \textbf{inherent variability in ground-truth masks} by
    leveraging multiple-annotator datasets
    \item We demonstrate an efficient \textbf{single forward pass} method to estimate both image and pixel-level uncertainties
    \item We offer insights into relevant downstream applications and conduct a comprehensive analysis of method components, addressing recommendations from prior studies \cite{kahl2024values}. The code is publicly available on github.com.
\end{itemize}

\begin{figure}[t!]
    \centering
    \includegraphics[width=\textwidth]{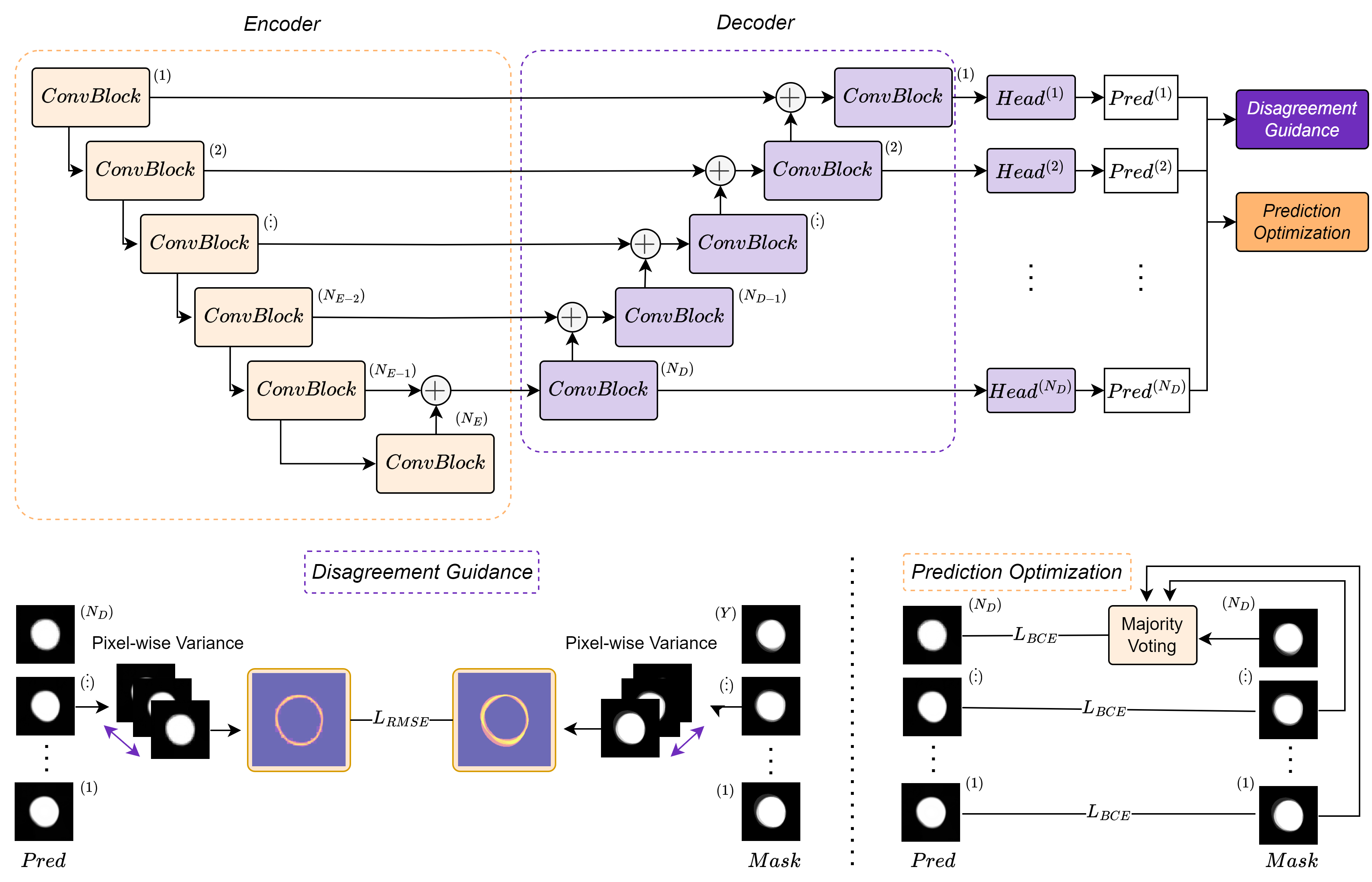}
    \caption{EDUE follows a U-Net-like architecture. The disagreement guidance module captures uncertainty by comparing variance heatmaps from the model and labels. Prediction optimization with a random sampling strategy is used for segmentation outputs.}
    \label{fig:network}
\end{figure}

\section{Methodology} 
We build upon Layer Ensembles (LE) \cite{kushibar2022layer}, a single-pass uncertainty estimation network for medical image segmentation. 
% They utilize the concept of prediction depth (PD) \cite{baldock2021deep}, which is used for assessing sample complexity.
The authors extended the concept of prediction depth (PD) \cite{baldock2021deep}, which is used for assessing sample complexity and segmentation, using segmentation heads after each encoder and decoder block for uncertainty estimation. Our goal is to link PD, multiple segmentation heads, and ground truth variability in multi-annotator datasets to enhance the UE capabilities of the models.

\begin{table}[t!]
    \caption{Results on the RIGA dataset. 
    The values are the mean (standard deviation) of three runs for LE and the proposed method. For DE, one run is used.}
    \label{tab:riga_dice_results}
    % \begin{minipage}{.9\textwidth}
        \centering
        \fontsize{8pt}{12pt}\selectfont
        \begin{tabular}{c|c|c|c|c|c|c|c|c|c}
            \hline
        \multicolumn{7}{c|}{Correlation Analysis} & \multicolumn{3}{c}{Segmentation Scores} \\
            \hline
            Method & \makecell{SR \\ Disc} $\uparrow$ & \makecell{SR \\ Cup} $\uparrow$ & \makecell{DC \\ Disc} $\uparrow$ & \makecell{DC \\ Cup} $\uparrow$ &  \makecell{NCC \\ Disc} $\uparrow$ & \makecell{NCC \\ Cup} $\uparrow$ & \makecell{Dice \\ Disc} $\uparrow$ & \makecell{Dice \\ Cup} $\uparrow$ & NLL $\downarrow$ \\
            \hline
            DE & 0.472 & 0.349 & 0.444 & 0.399 & \makecell{0.644} & \makecell{0.607} & \makecell{\textbf{0.977}} & \makecell{\textbf{0.874}} & \makecell{0.219} \\
            \hline
            LE & \makecell{-0.630 \\ (0.027)} & \makecell{-0.517 \\ (0.029)} & \makecell{0.601 \\ (0.025)} & \makecell{0.516 \\ (0.023)} & \makecell{0.470 \\ (0.059)} & \makecell{0.473 \\ (0.046)} & \makecell{0.974 \\ (0.001)} & \makecell{0.869 \\ (0.004)} & \makecell{0.174 \\ (0.001)} \\
            \hline
            Our & \makecell{\textbf{0.703} \\ (0.016)} & \makecell{\textbf{0.689} \\ (0.012)} & \makecell{\textbf{0.685} \\ (0.020)} & \makecell{\textbf{0.651} \\ (0.008)} & \makecell{\textbf{0.766} \\ (0.005)} & \makecell{\textbf{0.735} \\ (0.003)} & \makecell{0.970 \\ (0.000)} & \makecell{0.856 \\ (0.001)} & \makecell{\textbf{0.163} \\ (0.005)} \\
            \hline
        \end{tabular}
    % \end{minipage}%
\end{table}
EDUE model (Figure \ref{fig:network}) is comprised of four main components: an encoder for image feature extraction, a decoder with segmentation heads attached after each level, and a disagreement guidance module (DGM) for uncertainty estimation and prediction optimization of segmentation outputs.

\noindent\textbf{Encoder.} Let $\mathbf{X}\in\mathbb{R}^{H\times W\times C}$ denote an input image, where $H$ and $W$ are the width and height, and $C$ is the number of channels of the image that pass through an encoder $E(\mathbf{X})$ with $N_E$ downsampling blocks. Feature maps after each block are used to feed into the following blocks and skip connections during upsampling. Furthermore, these outputs are passed to the decoder block.  

\noindent\textbf{Decoder.} Next, the outputs of each decoder block $\mathbf{F}_i\in\mathbb{R}^{C^{\prime}_i\times H^{\prime}_i W^{\prime}_i}$, where $C^\prime_i, H^\prime_i, W^\prime_i$ are respective feature map dimensions and $i \in \{1, \ldots, N_D\}$, $N_D$ is the number of decoder blocks, are fed into their corresponding segmentation heads. The decoder blocks contain upsampling, convolutional layers, batch normalization, ReLU activation, and skip connections.

\noindent\textbf{DGM module.} The main component of the model is the DGM, which captures the variability in the ground truth masks from different clinicians. The bottom left part of Figure~\ref{fig:network} shows the module in detail. We stack $N_D$ prediction masks $\mathbf{\hat{M}}_i$, $i \in \{1, \ldots, N_D\}$, where $N_D$ is the total number of segmentation heads (attached after each decoder level), and compute pixel-wise variance along the channel axis to generate model uncertainty heatmap $\mathbf{\hat{H}}$. We repeat the same procedure for $Y$ ground truth masks $\mathbf{M}_j$, $j \in \{1, \ldots, Y\}$, where $Y$ is the number of annotations available for the given image, and we obtain the ground-truth variance heatmap $\mathbf{H}$. Next, we use the RMSE loss between two heatmaps to allow the model to learn the inherent uncertainty in the ground truth masks. In this way, each segmentation head can mimic a certain level of expertise, e.g., low, mid, and high-level details, based on the previously defined PD concept. Using heatmaps from ground-truth masks, these heads can imitate a high level of disagreement at the pixel level when the sample has many ambiguous regions and, on the other hand, have smaller uncertainty when the image is relatively easier to segment.

\noindent\textbf{Prediction Optimization.} Segmentation outputs are optimized in this block using a separate loss function. Previous studies \cite{jensen2019improving,jungo2018effect} indicate that networks trained with the ground truth from a single annotator tend to display overconfidence in predictions. Using multiple-annotator training schemes mitigates this issue and improves the model calibration. For each segmentation head, except the last, we randomly sample one of the annotations. For the last head, we use a soft majority voting label of the available $Y$ masks for the image. This block handles primary segmentation predictions, whereas the DGM captures uncertainty. Overall, the loss function can be defined as follows:

\begin{equation}
L = \alpha \cdot \sum_{i=1}^{N_D} L_{BCE_i}(\mathbf{\hat{M}}_{i}, \mathbf{M}_{i}) + \beta \cdot L_{RMSE}(\mathbf{\hat{H}}, \mathbf{H})
\end{equation}

where $L$ is the total loss, $L_{BCE_i}$ is the segmentation loss, $M_i$ are randomly sampled labels, except for the last head that uses soft majority voting label, $L_{RMSE}$ is used for modeling uncertainties, and $\alpha$, $\beta$ are balancing coefficients. 

\begin{table}[t]
    \caption{Results on the HECKTOR dataset for DE, LE, and proposed method. The values displayed are the mean (standard deviation) of 3-fold cross-validation.}
    \label{tab:hecktor_dice_results}
    % \begin{minipage}{.9\textwidth}
        \centering
        \fontsize{8pt}{12pt}\selectfont
        \begin{tabular}{c|c|c|c|c|c|c|c|c|c}
            \hline
        \multicolumn{7}{c|}{Correlation Analysis} & \multicolumn{3}{c}{Segmentation Scores} \\
            \hline
            Method & \makecell{SR \\ GTVp} $\uparrow$ & \makecell{SR \\ GTVn} $\uparrow$ & \makecell{DC \\ GTVp} $\uparrow$ & \makecell{DC \\ GTVn} $\uparrow$ &  \makecell{NCC \\ GTVp} $\uparrow$ & \makecell{NCC \\ GTVn} $\uparrow$ & \makecell{Dice \\ GTVp} $\uparrow$ & \makecell{Dice \\ GTVn} $\uparrow$ & \makecell{NLL\\$\times10^2$} $\downarrow$ \\
            \hline
            DE & \makecell{0.184 \\ (0.495)} & \makecell{0.158 \\(0.324)} & \makecell{0.349 \\(0.105)} & \makecell{0.245 \\(0.107)} & \makecell{{0.314} \\(0.025)} & \makecell{{0.348} \\(0.094)} & \makecell{\textbf{0.829} \\(0.062)} & \makecell{\textbf{0.784} \\(0.040)} & \makecell{0.512 \\(0.070)} \\
            \hline
            LE & \makecell{0.810 \\ (0.033)} & \makecell{0.620 \\ (0.049)} & \makecell{0.766 \\ (0.046)} & \makecell{0.575 \\ (0.063)} & \makecell{0.242 \\ (0.008)} & \makecell{0.245 \\ (0.100)} & \makecell{0.766 \\ (0.068)} & \makecell{0.691 \\ (0.055)} & \makecell{0.197 \\ (0.124)} \\
            \hline
            Our & \makecell{\textbf{0.904} \\ (0.015)} & \makecell{\textbf{0.816} \\ (0.085)} & \makecell{\textbf{0.862} \\ (0.071)} & \makecell{\textbf{0.779} \\ (0.057)} & \makecell{\textbf{0.456} \\ (0.083)} & \makecell{\textbf{0.492} \\ (0.083)} & \makecell{0.788 \\ (0.017)} & \makecell{0.726 \\ (0.003)} & \makecell{\textbf{0.131} \\ (0.097)} \\
            \hline
        \end{tabular}
    % \end{minipage}%
\end{table}

\section{Experimental Details}

\textbf{Datasets.} \textbf{RIGA} benchmark \cite{Almazroa2017-cp} is a public dataset for retinal cup and disc segmentation, containing 750 color fundus images from three databases: 460 from MESSIDOR, 195 from BinRushed, and 95 from Magrabia. Six glaucoma experts manually labeled the segmentation masks. For training, we used 195 samples from BinRushed and 460 from MESSIDOR, with the Magrabia dataset (95 samples) reserved for testing, as in prior works \cite{Liao2023TAB,ji2021learning}. All images are normalized between 0 and 1 and resized to 256$\times$256.

\noindent A subset of \textbf{HECKTOR 2022} data \cite{ANDREARCZYK2023102972} consists of 44 cases with multiple annotations, each with 3D CT, 3D PET of head and neck (H$\&$N) region rigidly registered to a standard frame, but at different resolutions. Annotations of gross tumor volumes of the primary tumors (GTVp) and lymph nodes (GTVn) come from 10 different experts. There are 3 annotations available for an image on average (ranges from 2 to 4 per patient). The annotations are private. 
We perform preprocessing steps similar to \cite{myronenko2022automated} and convert 3D volumes to 2D axial slices.  

\noindent\textbf{Comparison and Evaluation Metrics.} 
We compare our method with state-of-the-art DE and LE \cite{kushibar2022layer}. We measure image-level uncertainty correlations using Spearman's rank correlation (SR) and distance correlation (DC) \cite{dc_10.1214/009053607000000505} between variance sums in heatmaps from models and ground-truth masks. For pixel-level assessment, we compute the average normalized cross-correlation (NCC) between heatmap pairs for the test set. The Negative Log-Likelihood metric (NLL) evaluates network confidence calibration, penalizing small uncertainty for incorrect predictions. Segmentation performance is assessed using the soft Dice metric with soft majority voting labels as ground-truth masks. For the RIGA dataset, we report the results of three runs for EDUE and LE and a single run for DE. With HECKTOR, we report patient-based 3-fold cross-validation results.

\noindent \textbf{Implementation Details.}
We use a U-Net architecture ($N_E = 6, N_D = 5$) with a ResNet50 pre-trained on ImageNet as the encoder. For the RIGA dataset, the models are trained for 200 epochs with a batch size of 16, and for HECKTOR, it's 120 epochs with a batch size of 32; chosen empirically. The learning rate remains constant at $5e^{-5}$. We employ 5 networks in DE. We skip the first 5 segmentation heads for LE as suggested by the original authors \cite{kushibar2022layer}. For our method, we set $\beta$ to 5 and 2.5 for the RIGA and HECKTOR datasets, respectively.

\begin{figure}[t!]
    \centering 
    \subfigure[]{\includegraphics[width=0.4\textwidth]{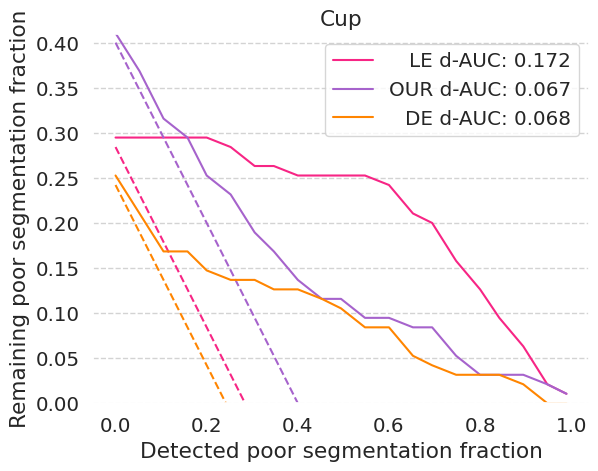}} 
    \subfigure[]{\includegraphics[width=0.4\textwidth]{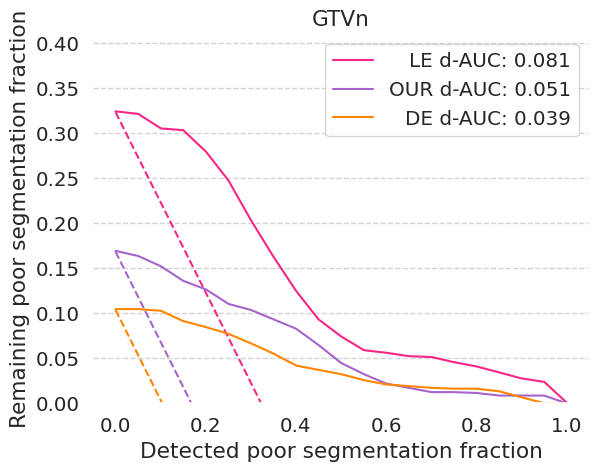}}
    \caption{Segmentation quality control results for LE, proposed method, and DE. Corresponding dashed lines are ideal lines. d-AUC: the difference between the area under the main curve and the ideal line (lower is better). (a) Cup results (b) GTVn results.}
    \label{fig:seg_qual_control}
\end{figure}

\section{Results and Discussion}

\begin{figure}[t!]
    \centering
    \subfigure[]{\includegraphics[width=0.4\textwidth]{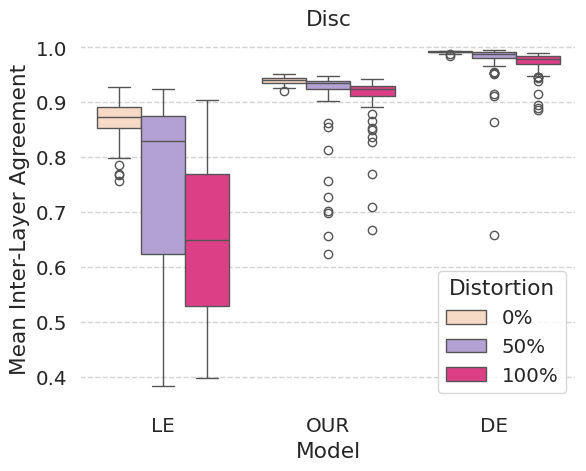}} 
    \subfigure[]{\includegraphics[width=0.4\textwidth]{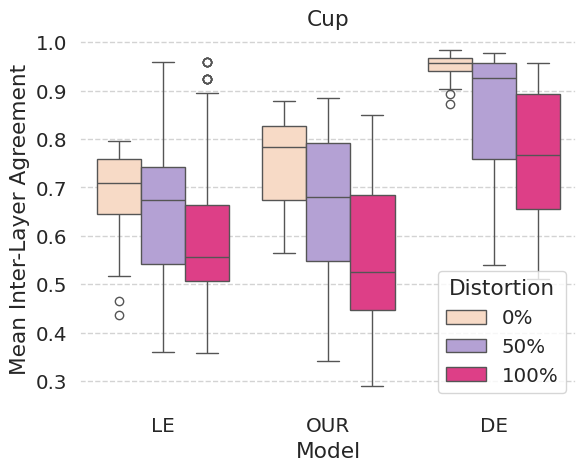}} 
    \caption{Out-of-distribution detection results. The box plots show the distribution of layer agreements and model agreements for LE, EDUE, and DE, respectively, at 0$\%$, 50$\%$, and 100$\%$ of distorted images. (a) Disc results (b) Cup results}
    \label{fig:ood_fig}
\end{figure}

\begin{figure}[b!]
    \center
    \includegraphics[width=0.8\textwidth]
    {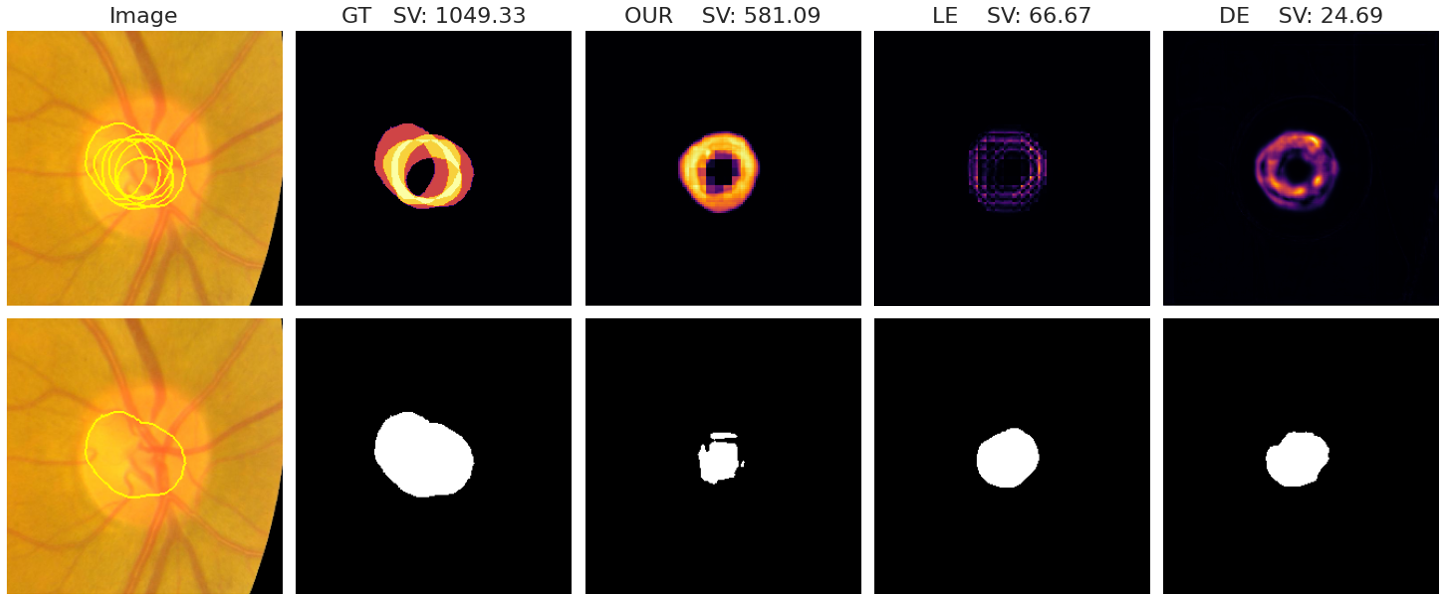}
    \label{fig:qual_riga_cup_img}
    \caption{Sample (cup) from the RIGA dataset. \textbf{Top row:} Input image with contours of all masks, ground-truth variance heatmap, variance heatmaps from EDUE, LE, DE. \textbf{Bottom row:} Input image with soft majority voting mask's contour (threshold 0.5), corresponding ground-truth mask, predicted masks from EDUE, LE, DE. SV: sum of variances.}
\end{figure}

\subsection{Correlation Analysis and Segmentation Performance}
Table~\ref{tab:riga_dice_results} and Table~\ref{tab:hecktor_dice_results} summarize the correlation and segmentation performance for all the methods for RIGA and HECKTOR datasets.
In Table~\ref{tab:riga_dice_results}, DE shows a moderate correlation with the SR value reaching around 0.47 and 0.35 for the disc and cup, respectively, whereas LE indicates a negative correlation. EDUE shows a stronger correlation with SR of 0.703 (0.02) and 0.689 (0.012) for the disc and cup respectively. This comparison is reflected in the DC scores as well. While DE and LE models show some correlation, our method shows much higher DC values for both the disc and cup, with 0.685 (0.020) and 0.651 (0.008). At the pixel level, NCC values for our model are also the highest, with 0.77 and 0.73 for disc and cup, respectively. EDUE has a slight drop in segmentation performance - 0.007 in the disc and 0.018 in the cup compared to the best-performing DE model. However, it has the lowest NLL value of 0.163 (0.005), while DE and LE models are more (or even over-) confident in their predictions. This suggests the proposed model is less prone to overconfidence and making predictions at somewhat ambiguous regions in the image. We confirm this behavior in the qualitative examples (Figure \ref{fig:qual_riga_cup_img} and Figure A2 in Appendix) and by examining the $\beta$ coefficient (Table A1 in Appendix). In addition, it is important to recall that DE has five times more model parameters compared to EDUE and LE.

The HECKTOR dataset shows a similar trend in performance, both in correlation and segmentation analyses, as listed in Table~\ref{tab:hecktor_dice_results}. The DE model captures a low level of correlation in the GTVp and GTVn in SR and DC metrics. While LE shows reasonable scores, EDUE achieves the highest correlation scores at both image and pixel levels. DE has the highest dice scores, however, LE and EDUE show much better calibration, with the latter achieving the lowest NLL.

\subsection{Segmentation Quality Control}

We evaluate methods for segmentation quality control \cite{ng2022estimating} on both datasets. We choose dice thresholds of 0.97 and 0.85 for disc and cup, 0.65 and 0.55 for GTVp and GTVn respectively to mark model outputs as poor segmentations. The variance metric is used for all methods to detect these marked masks. Figure \ref{fig:seg_qual_control} shows the results for cup and GTVn structures, which are harder tasks, while the results for disc and GTVp are available in Figure A1 in Appendix.
DE shows the best performance overall, with the lowest remaining poor segmentation fraction at all quantile thresholds for both cup and GTVn. EDUE performs similarly to DE, especially for the cup. However, it has a slightly higher remaining poor segmentation fraction than DE for GTVn at higher variance thresholds. LE has the worst performance overall, with the highest remaining poor segmentation fractions. Overall, the results show that EDUE is effective for segmentation quality control, performing better than LE and similarly to DE.

\subsection{Out-of-Distribution Detection}
We also evaluate the methods for out-of-distribution detection (Figure \ref{fig:ood_fig}) by applying Gaussian noise, blurring, hue, saturation, and value changes on the RIGA dataset. We use inter-layer agreement scores for EDUE, LE, and agreement scores between models for DE to detect outliers. Overall, the agreement scores decrease for all methods as the distortion level (proportion of images with augmentations) increases. This indicates that all methods are effective at detecting out-of-distribution samples as the applied distortions become more severe, which is more apparent with the cup as it has a more complex structure.

% \subsection{Effect of Heatmap Loss} -> appendix, mention somewhere

% \section{Discussion}
% combiner idea, the concept of explicitly assessing ucnertainty estimates can be generalized to other tasks and fields, with transformers, frameworks or strategies for choosing hard sampels and asking experts to labels them to have a subset that can be used for testing, doing pretraining in such styles, having similart pretext/self-supervised tasks?
% Overall approach is architecture agnostic and can be extended to other networks

\section{Conclusion}
We proposed a novel and simple UE approach that leverages expert disagreements to guide the model during training for improved uncertainty quantification capabilities without additional costs. Our extensive evaluations on the RIGA and HECKTOR datasets have shown that EDUE not only correlates strongly with expert opinions but also maintains robust segmentation performance. The alignment of model uncertainty with expert variability is a significant step towards fostering trust and transparency in clinical settings, which is crucial for the adoption of deep learning models in medical practice. 

% We think the overall idea is generalizable and opens doors for 

% Though simple, DGM can effectively improve model calibration ahd its uncertainty estimation capacities, without additional costs.

\bibliographystyle{splncs04}
\bibliography{bib}

\begin{thebibliography}{10}
\providecommand{\url}[1]{\texttt{#1}}
\providecommand{\urlprefix}{URL }
\providecommand{\doi}[1]{https://doi.org/#1}

\bibitem{Almazroa2017-cp}
Almazroa, A., Alodhayb, S., Osman, E., Ramadan, E., Hummadi, M., Dlaim, M., Alkatee, M., Raahemifar, K., Lakshminarayanan, V.: Agreement among ophthalmologists in marking the optic disc and optic cup in fundus images. Int. Ophthalmol.  \textbf{37}(3),  701--717 (Jun 2017)

\bibitem{ANDREARCZYK2023102972}
Andrearczyk, V., Oreiller, V., Boughdad, S., {Le Rest}, C.C., Tankyevych, O., Elhalawani, H., Jreige, M., Prior, J.O., Vallières, M., Visvikis, D., Hatt, M., Depeursinge, A.: Automatic head and neck tumor segmentation and outcome prediction relying on fdg-pet/ct images: Findings from the second edition of the hecktor challenge. Medical Image Analysis  \textbf{90},  102972 (2023). \doi{https://doi.org/10.1016/j.media.2023.102972}, \url{https://www.sciencedirect.com/science/article/pii/S1361841523002323}

\bibitem{Ashukha2020PitfallsOI}
Ashukha, A., Lyzhov, A., Molchanov, D., Vetrov, D.P.: Pitfalls of in-domain uncertainty estimation and ensembling in deep learning. ArXiv  \textbf{abs/2002.06470} (2020), \url{https://api.semanticscholar.org/CorpusID:209314627}

\bibitem{Ayhan2018TesttimeDA}
Ayhan, M.S., Berens, P.: Test-time data augmentation for estimation of heteroscedastic aleatoric uncertainty in deep neural networks (2018), \url{https://api.semanticscholar.org/CorpusID:13998356}

\bibitem{baldock2021deep}
Baldock, R.J.N., Maennel, H., Neyshabur, B.: Deep learning through the lens of example difficulty. Advances in Neural Information Processing Systems  \textbf{34} (2021)

\bibitem{banerji2023clinical}
Banerji, C.R., Chakraborti, T., Harbron, C., MacArthur, B.D.: Clinical ai tools must convey predictive uncertainty for each individual patient. Nature medicine  \textbf{29}(12),  2996--2998 (2023)

\bibitem{Blundell2015WeightUI}
Blundell, C., Cornebise, J., Kavukcuoglu, K., Wierstra, D.: Weight uncertainty in neural network. In: International Conference on Machine Learning (2015), \url{https://api.semanticscholar.org/CorpusID:39895556}

\bibitem{pmlr-v48-gal16}
Gal, Y., Ghahramani, Z.: Dropout as a bayesian approximation: Representing model uncertainty in deep learning. In: Balcan, M.F., Weinberger, K.Q. (eds.) Proceedings of The 33rd International Conference on Machine Learning. Proceedings of Machine Learning Research, vol.~48, pp. 1050--1059. PMLR, New York, New York, USA (20--22 Jun 2016), \url{https://proceedings.mlr.press/v48/gal16.html}

\bibitem{Gawlikowski2023}
Gawlikowski, J., Tassi, C.R.N., Ali, M., Lee, J., Humt, M., Feng, J., Kruspe, A., Triebel, R., Jung, P., Roscher, R., Shahzad, M., Yang, W., Bamler, R., Zhu, X.X.: A survey of uncertainty in deep neural networks. Artificial Intelligence Review  \textbf{56}(1),  1513--1589 (Oct 2023). \doi{10.1007/s10462-023-10562-9}, \url{https://doi.org/10.1007/s10462-023-10562-9}

\bibitem{jensen2019improving}
Jensen, M.H., J{\o}rgensen, D.R., Jalaboi, R., Hansen, M.E., Olsen, M.A.: Improving uncertainty estimation in convolutional neural networks using inter-rater agreement. In: Medical Image Computing and Computer Assisted Intervention--MICCAI 2019: 22nd International Conference, Shenzhen, China, October 13--17, 2019, Proceedings, Part IV 22. pp. 540--548. Springer (2019)

\bibitem{ji2021learning}
Ji, W., Yu, S., Wu, J., Ma, K., Bian, C., Bi, Q., Li, J., Liu, H., Cheng, L., Zheng, Y.: Learning calibrated medical image segmentation via multi-rater agreement modeling. Proceedings of the IEEE/CVF Conference on Computer Vision and Pattern Recognition (CVPR)  \textbf{3}(1),  12341--12351 (2021). \doi{10.1109/CVPR.2021.01241}

\bibitem{joskowicz2019inter}
Joskowicz, L., Cohen, D., Caplan, N., Sosna, J.: Inter-observer variability of manual contour delineation of structures in ct. European Radiology  \textbf{29}(3),  1391--1399 (2019). \doi{10.1007/s00330-018-5695-5}

\bibitem{jungo2018effect}
Jungo, A., Meier, R., Ermis, E., Blatti-Moreno, M., Herrmann, E., Wiest, R., Reyes, M.: On the effect of inter-observer variability for a reliable estimation of uncertainty of medical image segmentation. In: Medical Image Computing and Computer Assisted Intervention--MICCAI 2018: 21st International Conference, Granada, Spain, September 16-20, 2018, Proceedings, Part I. pp. 682--690. Springer (2018)

\bibitem{kahl2024values}
Kahl, K.C., L{\"u}th, C.T., Zenk, M., Maier-Hein, K., Jaeger, P.F.: Values: A framework for systematic validation of uncertainty estimation in semantic segmentation. arXiv preprint arXiv:2401.08501  (2024)

\bibitem{kumar2007glaucoma}
Kumar, S., Giubilato, A., Morgan, W., Jitskaia, L., Barry, C., Bulsara, M., Constable, I.J., Yogesan, K.: Glaucoma screening: analysis of conventional and telemedicine-friendly devices. Clinical \& Experimental Ophthalmology  \textbf{35}(3),  237--243 (2007). \doi{10.1111/j.1442-9071.2007.01457.x}

\bibitem{kushibar2022layer}
Kushibar, K., Campello, V., Garrucho, L., Linardos, A., Radeva, P., Lekadir, K.: Layer ensembles: A single-pass uncertainty estimation in deep learning for segmentation. In: International Conference on Medical Image Computing and Computer-Assisted Intervention. pp. 514--524. Springer (2022)

\bibitem{NIPS2017_9ef2ed4b}
Lakshminarayanan, B., Pritzel, A., Blundell, C.: Simple and scalable predictive uncertainty estimation using deep ensembles. In: Guyon, I., Luxburg, U.V., Bengio, S., Wallach, H., Fergus, R., Vishwanathan, S., Garnett, R. (eds.) Advances in Neural Information Processing Systems. vol.~30. Curran Associates, Inc. (2017), \url{https://proceedings.neurips.cc/paper_files/paper/2017/file/9ef2ed4b7fd2c810847ffa5fa85bce38-Paper.pdf}

\bibitem{Liao2023TAB}
Liao, Z., Hu, S., Xie, Y., Xia, Y.: Transformer-based annotation bias-aware medical image segmentation. In: International Conference on Medical Image Computing and Computer-Assisted Intervention. Springer (2023)

\bibitem{Mehrtens2023BenchmarkingCU}
Mehrtens, H.A., Kurz, A., Bucher, T.C., Brinker, T.J.: Benchmarking common uncertainty estimation methods with histopathological images under domain shift and label noise. Medical image analysis  \textbf{89},  102914 (2023), \url{https://api.semanticscholar.org/CorpusID:255393751}

\bibitem{myronenko2022automated}
Myronenko, A., Siddiquee, M.M.R., Yang, D., He, Y., Xu, D.: Automated head and neck tumor segmentation from 3d pet/ct (2022)

\bibitem{ng2022estimating}
Ng, M., Guo, F., Biswas, L., Petersen, S.E., Piechnik, S.K., Neubauer, S., Wright, G.: Estimating uncertainty in neural networks for cardiac mri segmentation: A benchmark study. IEEE Transactions on Biomedical Engineering  \textbf{69}(1),  1--23 (2022). \doi{10.1109/TBME.2022.3232730}

\bibitem{schaekermann2019understanding}
Schaekermann, M., Beaton, G., Habib, M., Lim, A., Larson, K., Law, E.: Understanding expert disagreement in medical data analysis through structured adjudication. Proceedings of the ACM on Human-Computer Interaction  \textbf{3}(1),  1--23 (2019). \doi{10.1145/3359178}

\bibitem{dc_10.1214/009053607000000505}
Sz{\'e}kely, G.J., Rizzo, M.L., Bakirov, N.K.: {Measuring and testing dependence by correlation of distances}. The Annals of Statistics  \textbf{35}(6),  2769 -- 2794 (2007). \doi{10.1214/009053607000000505}, \url{https://doi.org/10.1214/009053607000000505}

\bibitem{ZOU2023100003}
Zou, K., Chen, Z., Yuan, X., Shen, X., Wang, M., Fu, H.: A review of uncertainty estimation and its application in medical imaging. Meta-Radiology  \textbf{1}(1),  100003 (2023). \doi{https://doi.org/10.1016/j.metrad.2023.100003}, \url{https://www.sciencedirect.com/science/article/pii/S2950162823000036}

\end{thebibliography}
\end{document}